\documentclass{kapproc}
\usepackage{procps}
\usepackage[dvips]{graphicx}

\def\l{\left}
\def\r{\right}
\def\gsim{\mathrel{\rlap{\lower0.2em\hbox{$\sim$}}\raise0.2em\hbox{$>$}}}
\def\SumInt{\hbox{$\sum$}\hspace{-1.1em}\int\,}

\let\footnote\savefootnote

\let\footnotetext\savefootnotetext

\normallatexbib 

\begin{document}
\articletitle{The QCD equation of state \\ and quark star properties}

\author{A.~Peshier\footnote{Talk given at
 {\em Superdense QCD Matter and Compact Stars},
 27 September - 4 October 2003,
 Yerivan, Armenia}}
\affil{Institut f{\"u}r Theoretische Physik,
       University Giessen, 35392 Giessen, Germany}
\email{Andre.Peshier@theo.physik.uni-giessen.de}

\author{B.~K{\"a}mpfer}
\affil{Forschungszentrum Rossendorf, PF 510119, 01314 Dresden, Germany}
\email{B.Kaempfer@fz-rossendorf.de}

\author{G.~Soff}
\affil{Institut f{\"u}r Theoretische Physik,
       TU Dresden, 01062 Dresden, Germany}
\email{soff@physik.tu-dresden.de}

\begin{abstract}
We review our quasiparticle model for the thermodynamics of strongly interacting matter at high temperature, and its extrapolation to non-zero chemical potential. Some implications of the resulting soft equation of state of quark matter at low temperatures are pointed out.
\end{abstract}

\begin{keywords}
QCD equation of state, quasiparticles, quark stars
\end{keywords}

\section{Introduction}
The question for the equation of state of strongly interacting matter is a link between many-particle physics and astrophysics/cosmology. Calculated by means of statistical quantum field theory, it serves as a necessary input, e.\,g., in models of the early universe, or in the context considered here, it determines the structure of stars.
Astrophysical observations, such as the mass and the radius of dense stars, may in turn also impose constraints on the equation of state of deconfined quark matter.

We will focus on static spherically symmetric stars which are described by the Tolman-Oppenheimer-Volkov equations. At low densities, up to a few times nuclear density $n_0$, matter consists of interacting hadrons. Theoretical models for this state have to start from various assumptions, as for the included states and their interactions. Naturally, the results for the hadronic equation of state become notably model dependent at densities exceeding approximately $2n_0$. This is reflected in uncertainties of the predictions for the shell structure of neutron stars, cf.~\cite{Weber}.
At some higher energy density, hadronic matter undergoes a phase transition or a crossover to a quark-gluon plasma (at high temperatures), or a color-superconducting state (at low temperatures). Then the system can be described directly in terms of the fundamental degrees of freedom -- quarks and gluons. Notwithstanding, the coupling strength is still large in the regime of physical interest, and perturbative QCD is not reliable or, at least, the calculations have to be interpreted very carefully. Calculations based on various effective theories, on the other hand, rest again on assumptions which seem hard to control a priori.

Non-perturbative results have been obtained from first principles by lattice QCD computations at zero chemical potential and temperatures up to a few times the transition temperature $T_c$.
At low temperatures and for non-zero chemical potential $\mu$, as relevant for dense stars, Monte-Carlo calculations are, however, hampered by the sign problem. Several approaches to cope with it have been proposed only recently.
The available results still have rather large uncertainties, and they do not yet cover the range of temperatures and chemical potential required in the present context.

This makes worthwhile an approach which extrapolates, with as few assumptions as possible, lattice QCD data from zero chemical potential to $\mu>0$.
In the following we will outline a thermodynamical quasiparticle model, which can be derived in a series of approximations.

\section{Resummation and quasiparticle models}

\subsection{$\phi^4$ theory}
For the sake of simplicity we consider, for the moment being, a macroscopic system described by a scalar field theory.
Following \cite{LuttiW}, the thermodynamic potential, $\Omega = -pV$, can be calculated from a functional of the full propagator $\Delta = (\Delta_0^{-1}-\Pi)^{-1}$,
\begin{equation}
  \Omega
  =
  \textstyle\frac12\displaystyle\,
  \SumInt\l( \ln(-\Delta^{\!-1}) + \Pi\Delta\rule{0em}{1.2em} \r)
  -\Phi[\Delta] \, ,
  \label{eq: Omega_phi4}
\end{equation}
evaluated at the stationary point $\delta\Omega/\delta\Delta = 0$. This corresponds to a self-consistent calculation of the self-energy, $\Pi = 2 \delta\Phi / \delta\Delta$, where $\Phi$ is the series of 2-particle irreducible diagrams.

In this scheme, thermodynamically self-consistent approximations \cite{Baym} can be derived by truncating the expansion of $\Phi$, which amounts to a resummation of whole classes of diagrams in perturbation theory.
To leading order, for the massless case and in the $\overline{MS}$ scheme in $4-2\epsilon$ dimensions,
\begin{eqnarray}
 \Phi
 &=&
 \hskip -2mm \raise-.4em\hbox{
 \includegraphics[scale=0.5]{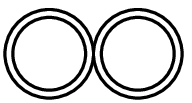}}
 \,=\,
 \sum
 \hskip -1mm \raise-5mm\hbox{
 \includegraphics[scale=0.25]{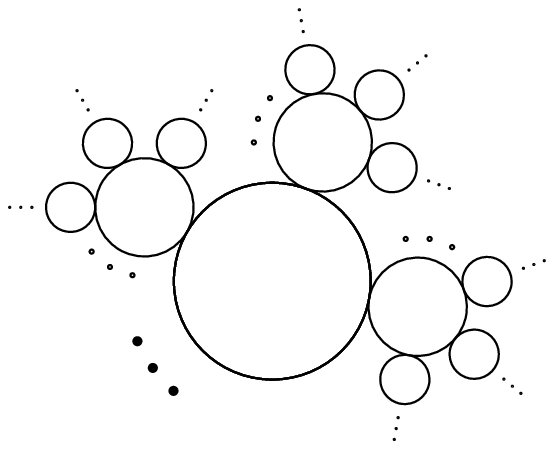}}
 \nonumber \\
 \Pi
 &=&
 \frac{-g_0^2}{4!}
 \l[
   \frac\Pi{16\pi^2}
    \l( \frac1\epsilon+\ln\frac{\bar\mu^2}\Pi+1 \r)
   -\int\frac{d^3k}{(2\pi)^3} \frac{n_b(\omega/T)}\omega
 \r] ,
\end{eqnarray}
where $n_b(x) = (e^x-1)^{-1}$ and $\omega = (k^2+\Pi)^{1/2}$, the self-energy is simply a mass term.
The non-perturbative gap equation requires the renormalization of the bare coupling $g_0$. The resummation of the set of `chain' graphs yields for the coupling at the scale of Mandelstam $s$
\begin{equation}
  g^2(s)
  =
  g_0^2
  -
  \frac{g_0^2}{4!}\,
  \frac3{4\pi^2} \l[ \frac1\epsilon +\ln\frac{\bar\mu^2}{-s}+2 \r]
  g^2(s) \, .
\end{equation}
Expressing now $g_0$ by $g(s)$ leads to the well-defined relation
\begin{equation}
 \Pi
 =
 \frac{g^2(s)}2
  \l[
   \frac\Pi{16\pi^2} \l( \ln\frac\Pi{-s} + 1 \r)
    +\int_{k^3} \frac{n_b(\omega/T)}\omega
 \r] ,
\end{equation}
whose solution is interpreted as a temperature dependent quasiparticle mass squared. The corresponding pressure reads
\begin{equation}
 p
 =
 -T \int_{k^3} \ln\l( 1-{\rm e}^{-\omega/T} \r)
 +\frac\Pi4 \int_{k^3} \frac{n_b(\omega/T)}\omega
 +\frac{\Pi^2}{128\pi^2} \, .
\end{equation}
For both calculational details as well as for a discussion of this approximation and its relation to other approaches we refer to \cite{PeshiKPS98}; here we only emphasize its structure. The first term on the rhs is simply the pressure of free massive particles. Written in the form $p=p^{\rm id}(T,m)-B(T)$, the function $B$ is related to $m(T)$ such that the entropy $s = \partial p/\partial T$ reduces, due to the stationarity of $\Omega$, to the entropy $s^{\rm id}(T,m)$ of an ideal gas.

\subsection{QCD at finite temperature}
\subsubsection{HTL quasiparticle model}
\begin{figure}[hbt]
 \centerline{\includegraphics[scale=0.55]{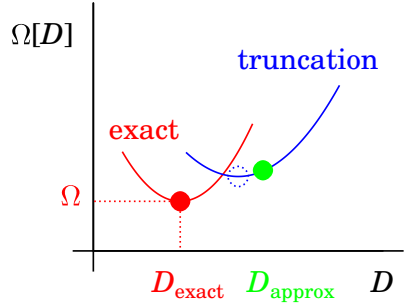}}
 \vskip -5mm
 \caption{The exact and the truncated functional for the thermodynamic
    potential. For the latter, a propagator `near' to the stationary
    point (which itself may be unphysical) can give a physically reasonable
    approximation.
    \label{fig: asc}}
\end{figure}

\begin{figure}[hbt]
 \centerline{\includegraphics[scale=0.75]{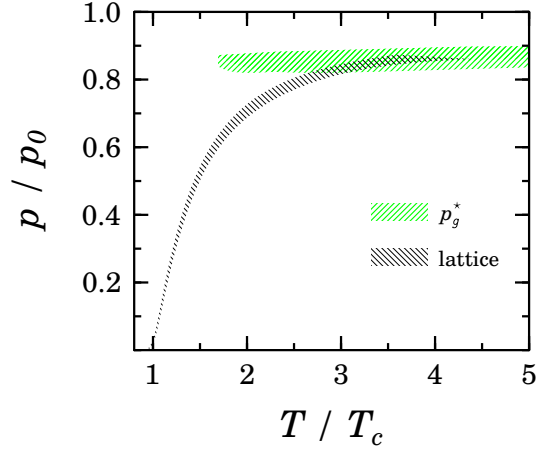}}
 \vskip -3mm
 \caption{Comparison of the pressure, in units of the free pressure,
    of the SU(3) plasma from the HTL quasiparticle approximation
    (\ref{eq: p*}) vs.\ lattice data \cite{lQCD}.
    \label{fig: p*}}
\end{figure}

In QCD, the truncation of a resummation scheme based on 2-point functions is delicate because of gauge invariance.
However, it can be argued that appropriate approximate solutions of the Schwinger-Dyson equations can yield reasonable approximations for $\Omega$, or the pressure, see the schematic Figure \ref{fig: asc}.
Putting this to a test, we consider the representative case of the pure gauge plasma,
\begin{equation}
 \includegraphics[scale=1]{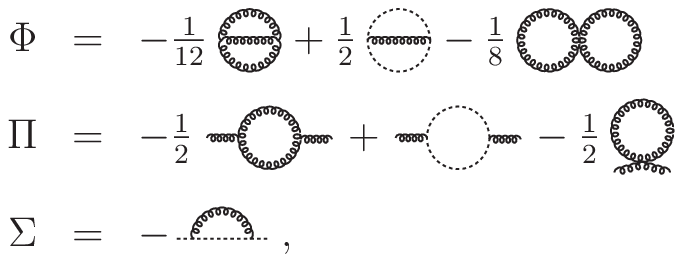}
\end{equation}
where the traces are taken over the group indices and the 4-momentum, and $\Sigma = G_0^{-1} - G^{-1}$ is the ghost self-energy. The propagators can be approximated by the hard thermal loop (HTL) contributions, which are gauge invariant and have the correct limit for hard momenta near the light cone. The resulting approximation for the pressure \cite{Peshi01}, in condensed form
\begin{equation}\textstyle
  p_g^\star
  =
 -\frac12\, {\rm Tr}
   \l[ \ln(-D_{\star}^{-1})+\frac12\,D_\star\Pi^\star \r]
  +{\rm Tr}[\ln(-G_0^{-1})] \, ,
  \label{eq: p*}
\end{equation}
has similar properties as the corresponding expression in the scalar theory. With the standard 2-loop running coupling, it agrees with the lattice data for temperatures down to $3T_c$, see Figure \ref{fig: p*},
which is a noteworthy improvement compared to the perturbative results.
Similar results have been obtained by calculating directly the HTL-resummed entropy \cite{BlaizIR01}.

\subsubsection{Phenomenological quasiparticle model}
\begin{figure}[hbt]
 \centerline{\includegraphics[scale=0.75]{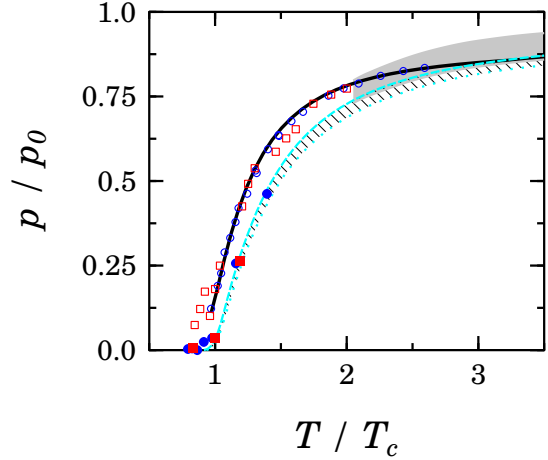}}
 \vskip -3mm
 \caption{The quasiparticle fit (solid line) of the lattice data
    \cite{CPPACS} (open symbols) for the pressure in QCD with $N_f=2$
    light flavors. The full symbols, representing data with large quark
    masses, agree with the results for the pure SU(3) plasma (hatched
    band); for details see \cite{PeshKS02}.
    \label{fig: p_eqp}}
\end{figure}

Taking into account only the dominant contributions in (\ref{eq: p*}), namely the quasiparticle contributions of the transverse gluons as well as the quark particle-excitations for $N_{\!f} \not= 0$, we arrive at the quasiparticle model \cite{PeshKPS96}. The dispersion relations can be even further simplified by their form at hard momenta, $\omega_i^2 = k^2+m_i^2$, where $m_i \sim gT$ are the asymptotic masses. With this approximation of the self-energies, the pressure reads in analogy to the scalar case
\begin{equation}
 p(T)
 =
 \sum_i p^{\rm id}(T, m_i) - B(T) \, .
 \label{eq: p_eqp}
\end{equation}
Conceding an enhancement of the running coupling in the infrared, parameterized by $T_s>0$ in an ansatz compatible with the perturbative limit,
\begin{equation}
   g^2(T)
   =
   \frac{48\pi^2}
   {(11N_c - 2\, N_{\!f})
     \ln\l( \frac{T-T_s}{T_c/\lambda} \r)^2} \, ,
\end{equation}
the thermodynamic lattice data can be quantitatively described even down to $T_c$, for an example see Figure \ref{fig: p_eqp}.
The coupling $g(T)$ obtained from available lattice QCD data with a given number of flavors will be an input for the extrapolation to non-zero chemical potential, as outlined in the following.

\subsection{Non-zero chemical potential}
\begin{figure}[hbt]
 \centerline{\includegraphics[scale=0.75]{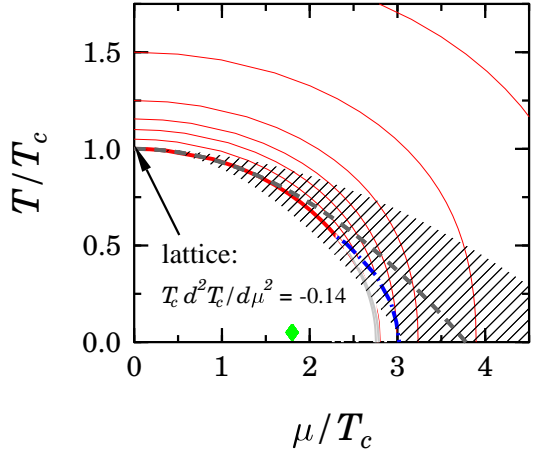}}
 \vskip -3mm
 \caption{The characteristics of the flow equation (\ref{eq: flow eq})
    for $N_f = 2$ light flavors. The innermost characteristic line
    coincides with the prediction from lattice QCD \cite{Allt} for the
    critical line at small $\mu$, which is represented by the dashed
    line with the hatched error band.
    \label{fig: char}}
\end{figure}

The phenomenological quasiparticle model can be generalized to non-zero chemical potential, where the quasiparticle masses of the gluons and quarks read
\begin{eqnarray}
  m_g^2
  &=&
  \frac16
  \l[
     \l( N_c+ \textstyle\frac12\, N_{\!f} \r) T^2
   + \frac3{2\pi^2} \sum_q \mu_q^2
  \r] g^2
  \nonumber \\
  m_q^2
  &=&
  \frac{N_c^2-1}{8N_c}\, \l[ T^2+\frac{\mu_q^2}{\pi^2} \r] g^2 \, ,
\end{eqnarray}
and the pressure, analogous to Equation (\ref{eq: p_eqp}), now depends also on $\mu$.
The predictive power of the approach becomes obvious by noting that the dependance of the coupling on $T$ and $\mu$ is completely governed by the requirement of thermodynamic consistency \cite{PeshKS00}: Maxwell's relation,
\[
 \frac{\partial s(T,\mu,m_i)}{\partial \mu}
 =
 \frac{\partial n(T,\mu,m_i)}{\partial T} \, ,
\]
where $n = \partial p / \partial \mu$ is the particle density, implies
\begin{equation}
  a_\mu(\mu,T,g^2) \frac{\partial g^2}{\partial \mu}
  +a_T(\mu,T,g^2) \frac{\partial g^2}{\partial T}
  =
  b(\mu,T,g^2) \, .
  \label{eq: flow eq}
\end{equation}
While nonlinear in $g^2$ (the coefficients $a_{\mu,T}$ and $b$ are lengthy integral expressions), the partial differential equation is linear in the derivatives. It can thus be solved by the method of characteristics, with the boundary conditions given by the coupling at $\mu = 0$, as obtained from finite-$T$ lattice QCD.
The resulting elliptic flow shown in Figure \ref{fig: char}, which maps the 
equation of state from $\mu = 0$ to $\mu > 0$, is plausible from the
physical intuition.
In the perturbative limit, $T \rightarrow c \mu/\pi$ with $c = (\frac{4N_c+5N_f}{9N_f})^{-1/4} \approx 1$.
It is noteworthy that the correspondence $T \sim \mu/\pi$ holds with a good accuracy even when $g$ is not small. A similar observation was made in the HTL quasiparticle approach \cite{Roma}.

\begin{figure}[hbt]
 \centerline{\includegraphics[scale=0.75]{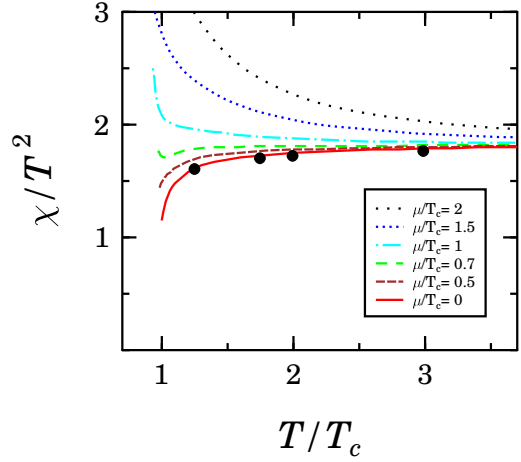}}
 \vskip -3mm
 \caption{The quark number susceptibility for $N_f=2$, calculated
    from the quasiparticle model with the same parameters as in
    Figs.~\ref{fig: p_eqp} and \ref{fig: char}, for several chemical
    potentials compared to the lattice data \cite{GavaGM} at $\mu=0$.
    \label{fig: chi_eqp}}
\end{figure}

The characteristic line emanating from $T_c$ is naturally related to the critical line $T_c(\mu)$ enclosing the hadronic phase. The comparison, in Figure \ref{fig: char}, of our result for the curvature of the critical line at $\mu=0$, which can be calculated in lattice QCD \cite{Allt}, is a nontrivial and successful test of the extension of the quasiparticle approach to $\mu>0$.
The quark number susceptibility $\chi(T) = \partial n / \partial\mu |_{\mu=0}$ is another quantity which has been computed on the lattice. As a second derivative of the pressure it is a very sensitive benchmark, and it agrees nicely with our result, see Figure \ref{fig: chi_eqp}.
Finally, as a direct confirmation of our mapping procedure, the quasiparticle model can also successfully describe the available lattice data for $p(\mu, T)$ with $2+1$ flavors \cite{SzabT}. It should be noted, however, that these direct calculations are so far restricted to small lattice sizes, resulting in still rather large uncertainties.
With these supporting arguments we consider our results from the extra\-polation of the lattice data at $\mu=0$ (with controllable small uncertainties) as a realistic estimate for the equation of state at not too small temperatures.

At low temperatures, matter will undergo a transition to a color-supercon\-ducting state, with a different quasiparticle structure than presumed in our quasiparticle approach. Nonetheless, pairing affects the thermodynamic bulk properties only at the relative order of ${\cal O}(\Delta^2/\mu^2)$, where the estimated gap $\Delta < 100\,$MeV is comfortably smaller than the chemical potential. Therefore, our equation of state is a reasonable approximation even at small temperatures (maybe except for the pressure where it becomes very small).

\begin{figure}[hbt]
 \centerline{\includegraphics[scale=0.75]{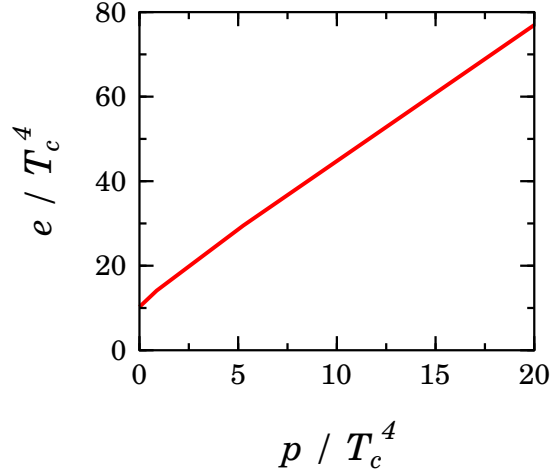}}
 \caption{The estimate for the equation of state, for $N_f=2$, at small
    temperatures.
    \label{fig: e(p)}}
\end{figure}

Relevant for the following discussion is the equation of state in the form $e(p)$, at $T \approx 0$. Although both the pressure and the energy density \begin{figure}[hbt]
 \includegraphics[scale=0.57,angle=-90]{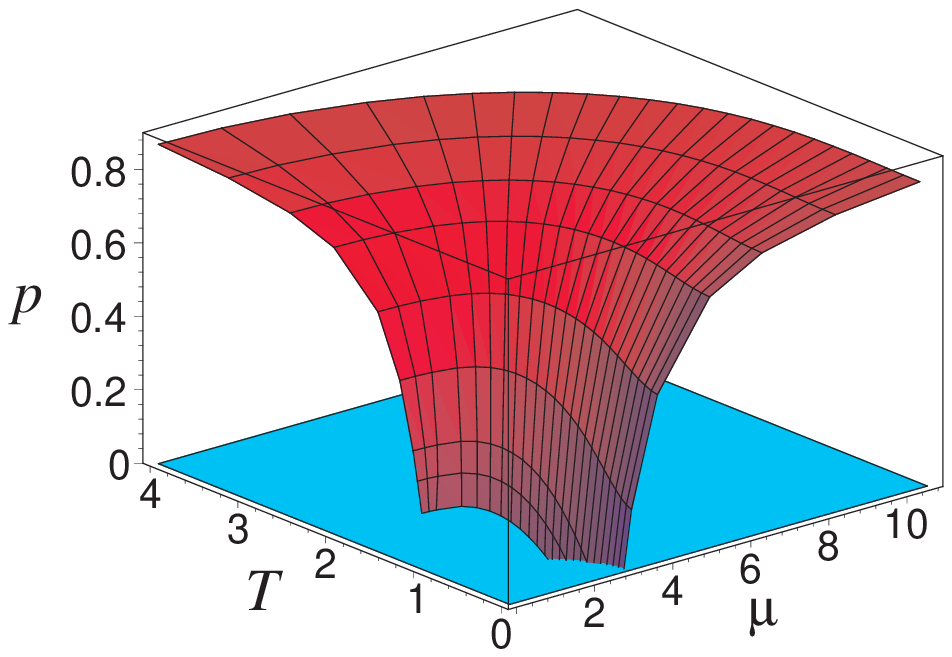}
 \hfill
 \raisebox{3mm}{
 \includegraphics[scale=0.57,angle=-90]{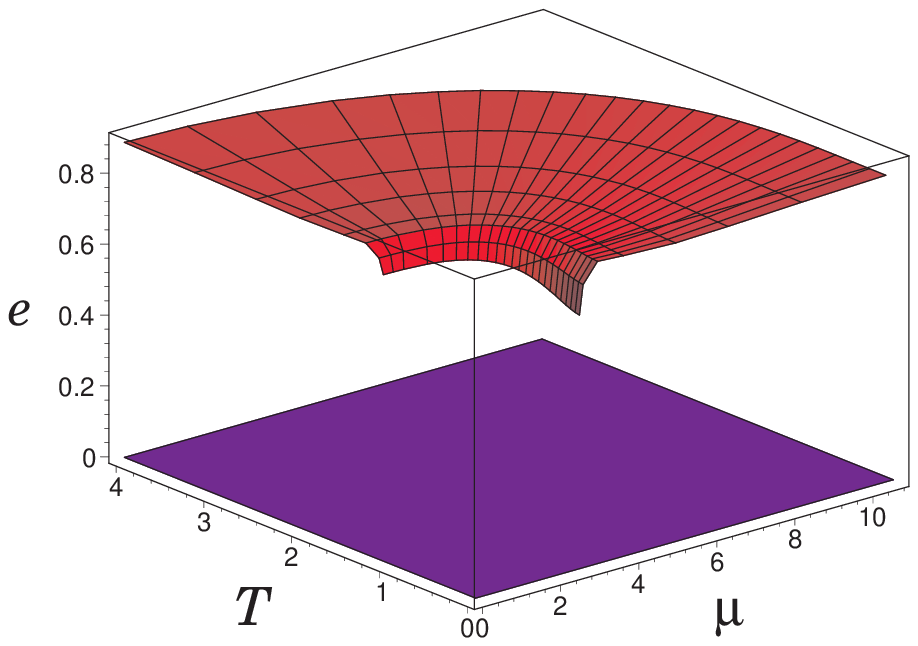}}
 \caption{The pressure and the energy density, scaled by the free
    results, for $N_f=2$.
    \label{fig: p&e (mu,T)}}
\end{figure}
deviate sizably from their ideal values, cf.~Figure \ref{fig: p&e (mu,T)},
we observe an almost linear relation
\begin{equation}
  e(p) = \alpha p + 4\tilde B \, ,
  \label{eq: e(p)}
\end{equation}
as shown in Figure \ref{fig: e(p)}.
The same linear form, with $\alpha = 3$, follows in the familiar MIT-bag approach where $p(x) = cx^4 - B_{\rm MIT}$ ($x$ is $\mu$ or $T$, and $c$ is some constant). Although this parameterization of the pressure is clearly in contradiction with the thermodynamic lattice results, the bag model relation for $e(p)$ serves as a standard to compare our results with.
We note in passing that a linear relation between $e$ and $p$, with a slope $\alpha \approx 3$ is to be expected since both quantities scale with the forth power of either $\mu$ or $T$. Causality requires furthermore $\alpha > 3$.
We find $\alpha \approx 3.2$, as a sign of the seen interaction effects, and $\tilde B \approx 3T_c^4$.

\section{Implications for quark stars}
Although there exist some lattice data for two light and one heavier quark, they still have larger uncertainties than for the case $N_{\!f}=2$ considered so far.
Nevertheless, based on the available data we can point out some interesting implications.
From the universality of the scaled pressure $p/p_0(T/T_c)$ for various numbers of flavors, as observed on the lattice \cite{KarsLP}, we can expect that also at $\mu > 0$ the results do not change much besides a scaling from the different number of degrees of freedom in the physical case.
On more general grounds we expect an approximately linear relation for $e(p)$ for large ranges of the pressure.

This expectation is confirmed by analyzing the available data, and
approximating the lepton component in $\beta$ equilibrated matter by an ideal gas.
The results for $e(p) \approx \alpha p + 4\tilde B$, in particular, are found to be rather insensitive even under large arbitrary variations of the values $\lambda$ and $T_s$ which parameterize $g(\mu=0,T)$.
The slope parameter is found to be constrained by $3 < \alpha < 3.5$.
The value of $\tilde B$, on the other hand, is in our approach directly linked to the transition temperature, $B \approx 3 T_c^4$. Since $T_c \approx 160\,$MeV is measured on the lattice, we have a definite prediction for the absolute scale in the equation of state.
Our estimate, $\tilde B^{1/4} \approx 210\,$MeV, is substantially larger than the typical value of the bag constant obtained from fitting hadron spectra, $B^{1/4}_{\rm MIT} \approx 150\,$MeV.

\begin{figure}[hbt]
 \centerline{\includegraphics[scale=0.75]{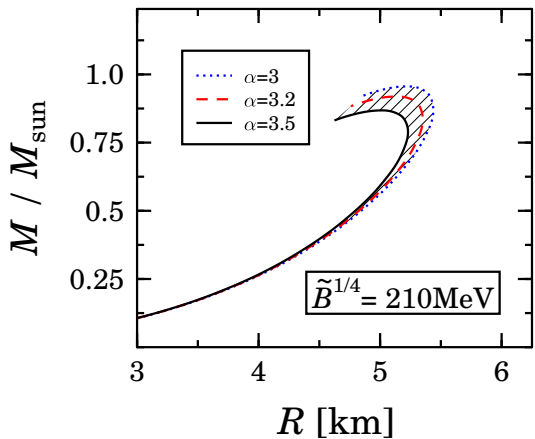}}
 \caption{The mass-radius relation, for the linear equation of state
    (\ref{eq: e(p)}), of a static quark star.
    \label{fig: MR rel}}
\end{figure}

For a linear equation of state, the Tolman-Oppenheimer-Volkov equations imply a scaling property for the total mass and the radius of the star,
\begin{equation}
  M, R \sim \tilde B^{-1/2} \, .
\end{equation}
The effects of deviations of $\alpha$ from 3 being small, cf.\ Figure \ref{fig: MR rel},
we pointed out \cite{PeshKS00} the possibility of the existence of very dense and compact objects, $M \approx 0.9M_{\rm sun}$ and $R \approx 6\,$km, composed mainly of quark matter.
Similar values for the maximal mass and radius were found in a perturbative approach with a physically motivated choice of the renormalization scale \cite{FragPS}. In a Schwinger-Dyson approach \cite{BlasGPRS}, $M \approx 0.7M_{\rm sun}$ and $R \approx 9\,$km were obtained.
It is interesting to compare these values with the mass and the radius of the object RX\,J1856: $M \approx 0.9M_{\rm sun}$ and $R \approx 6\,$km \cite{DrakEtAl}, which are not compatible with any hadronic equation of state. The precise values for $M$ and $R$ are, however, still under debate; for a recent discussion, we refer to \cite{ThomTB}.

\begin{figure}[hbt]
 \centerline{\includegraphics[scale=0.4]{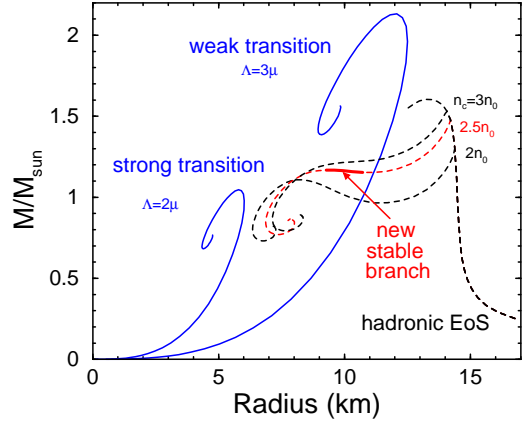}}
 \caption{For a strong first order transition, a new stable branch can
    exist in the mass-radius relation (Figure from \cite{FragPS}).
    \label{fig: new branch}}
\end{figure}

Taking into account the effects of the hadronic crust of the star, its properties become sensitive to details of the hadronic equation of state, and of the transition.
For a strong first order transition, as suggested in \cite{GerlKamp},
a new branch in the mass-radius diagram could exist, see Figure \ref{fig: new branch}.
The observation of so-called twins, as emphasized in \cite{ScheGST},
could be an exciting astrophysical indication of the existence of a non-hadronic phase in the center of neutron stars.
\\[1cm]
{\bf Acknowledgements} \quad
The work of A.~P.\ is supported by BMBF.

\end{document}